\documentclass[twocolumn,pra,aps,showpacs]{revtex4}
\usepackage{amsmath}
\usepackage{epsfig}

\begin{document}

\title{Quantum Entanglement Manifestation of Transition to Nonlinear
Self-trapping for Bose-Einstein Condensates in a Symmetric Double-Well}

\begin{abstract}
We investigate the nonlinear self-trapping phenomenon of the Bose-Einstein
condensates (BEC) in a symmetric double-well, emphasizing on its behind
dynamical phase transition. With increasing the nonlinear parameter
depicting the interaction between the degenerate atoms the BEC turns to be
self-trapped manifesting an asymmetric distribution of the atomic density
profile. Essence of this phenomenon is revealed to be a continuous phase
transition and underlying critical behavior is studied analytically and
found to follow a logarithm scaling-law. We then go beyond the mean field
treatment and extend to discuss the effect of the many-body quantum
fluctuation on the transition. It is found that the transition point is
shifted and the scaling-law is broken. In particular, the quantum phase
transition is accompanied by the change of the entanglement entropy which is
found to reach maximum at transition point. Behind physics is revealed.
\end{abstract}

\author{Li-Bin Fu and Jie Liu$^{*}$}
\affiliation{Institute of Applied Physics and Computational Mathematics, P.O. Box 8009
(28), 100088 Beijing, China}
\maketitle


\section{Introduction}

Double-well system is a paradigm model used to demonstrate marvellous
quantum tunnelling\cite{grifoni}. The realization of dilute Bose degenerate
gas in last nineties provides a possibility of directly observing the
tunnelling in the matter wave of macroscopic scale up to 100$\mu m$ \cite%
{anderson}. In Bose-Einstein condensates (BECs) system the interaction
between the degenerate ultra-cold atoms plays an crucial role. It
dramatically affects the quantum dynamics and leads to many unusual
phenomena like nonlinear Josephson oscillation, nonlinear quantum tunnelling
and critical onset in coherent oscillations , etc.\cite%
{franco,milburn,JieLiu,gio}. These problems have attracted much theoretical
attention over the past few years and the recent realization of the BECs in
the optical trap of a double-well configuration has brought a new research
surge\cite{shin,michael}.

Among many findings, the transition to self-trapping is most interesting one %
\cite{selftrap,smerzi,raghavan,raghavan1,sigmund}. It says, with increasing
the atomic interaction (repulsive), the Josephson oscillation between two
wells is blocked, the BECs atoms in a symmetric double-well potential shows
highly asymmetric distribution as if most atoms are trapped in one well.
This phenomenon was observed in lab recently\cite{michael}.

In this paper we achieve insight into this somehow counterintuitive
phenomenon with addressing its behind phase transition and the influence of
many-body quantum fluctuation on the phase transition. Analytically we
identify the self-trapping phenomenon as a continuous phase transition and
critical behavior is found to be characterized by a logarithm scaling-law.
We then extend to discuss many-body quantum fluctuation effect on
self-trapping. We find that, the transition point is shifted and the
scaling-law is broken down due to the quantum fluctuation. Further
investigations show that quantum properties are quite different for
different interaction regions, such as self-trapping region and others.
These properties can be well illustrated by entanglement entropy
quantitatively. With employing average entropy as the order parameter, we
can clearly demonstrate the quantum phase transition: The entanglement
entropy reaches its maximum at quantum transition point.

Our paper is organized as follows. In Sec.II, the self-trapping is studied
within mean-field framework both analytically and numerically, revealing the
critical behavior at the transition point. In Sec.III, we discuss the
many-body quantum fluctuation effect and reveal the quantum entanglement
manifestation of transition to nonlinear self-trapping. Sec.IV is our
discussion and conclusion.

\section{Transition to Self-trapping and scaling law}

For two weakly coupled BECs trapped in a symmetric double-well, the system
can be described by the so-called two-mode Hamiltonian \cite{franco,s14}
\begin{equation}
\widehat{H}=\frac{\gamma }{2}\left( \widehat{a}^{\dagger }\widehat{a}-%
\widehat{b}^{\dagger }\widehat{b}\right) +\frac{c}{2N}\left( \widehat{a}%
^{\dagger }\widehat{a}-\widehat{b}^{\dagger }\widehat{b}\right) ^{2}-\frac{v%
}{2}\left( \widehat{a}^{\dagger }\widehat{b}+\widehat{b}^{\dagger }\widehat{a%
}\right) ,  \label{ham}
\end{equation}
where the Bose operators $\widehat{a}^{(\dagger )}$ and $\widehat{b}%
^{(\dagger )}$ correspond to annihilating (creating) operators for different
well respectively, $\gamma =E_{a}^{0}-E_{b}^{0}$ is the energy bias between
the two wells and $E_{i}^{0}=\int \left( \frac{\hbar ^{2}}{2m}|\nabla
\varphi _{i}|^{2}+V(r)|\varphi _{i}|^{2}\right) dr$ , $c=c_{i}=\frac{4\pi
\hbar aN}{m}\int |\varphi _{i}|^{4}dr$ denotes the effective interaction of
atoms, $v=\int \left( \frac{\hbar ^{2}}{2m}\nabla \varphi _{a}\nabla \varphi
_{b}+V(r)\varphi _{a}\varphi _{b}\right) dr$ is the effective Rabi frequency
which describes the coupling between two wells, $N$ is total atoms number
which is conserved, $a$ is the s-wave scattering length, and $\varphi _{i}$ (%
$i=a,b$) are wave functions for each well respectively. In the present work
we focus on the case which has been realized in lab recently. For this case,
the potential is symmetric so that $\gamma =0$, and the interaction is
repulsive, i.e., $c>0.$

If the particle number is larger enough, the system can be well described in
the mean-field approximation. Under mean-field approximation, the dynamics
of the system is described by a classical Hamiltonian $H=\left\langle \Psi
_{GP}\left| \widehat{H}\right| \Psi _{GP}\right\rangle /N$ (up to a trivial
constant) in which $\left| \Psi _{GP}\right\rangle =\frac{1}{\sqrt{N!}}%
\left( a\widehat{a}^{\dagger }+b\widehat{b}^{\dagger }\right) ^{N}\left|
0\right\rangle $ is collective state of $N$-particle system \cite{franco,s14}%
. $a=\left| a\right| e^{i\theta _{a}}$ and $b=\left| b\right| e^{i\theta
_{b}}\ $are two $c$-numbers which correspond to the probability amplitudes
of atoms in two wells respectively. By introducing the population difference
$s=\left| b\right| ^{2}-\left| a\right| ^{2}$ and the relative phase $\theta
=\theta _{b}-\theta _{a},$ the classical Hamiltonian can be reduced to
\begin{equation}
H=-\frac{c}{2}s^{2}+v\sqrt{1-s^{2}}\cos \theta ,  \label{equation3}
\end{equation}
where $s$ and $\theta $ are canonical conjugate coordinates. Their equations
of motions are
\begin{equation}
\dot{s}=v\sqrt{1-s^2}\sin\theta,\dot{\theta}=-cs-\frac{vs}{\sqrt{1-s^2}}%
\cos\theta.  \label{eqq}
\end{equation}

Self-trapping motion refers to the trajectories whose average population
difference is not zero $<s>\neq 0$. In the experiment \cite{michael}, all
the atoms are placed initially in one well, i.e., $s(0)=1$ or $-1$. As
observed, with small interaction, the Josephson oscillation will be
observed, and with larger interaction, the self-trapping emerges. This
phenomenon can be well understood by the above classical Hamiltonian systems
(\ref{equation3}). Fig.1 plots the evolution of population difference $s$
and its average for different interactions calculated by (\ref{eqq}) with
initial condition $s(0)=1$. For $c/v$ is smaller than $2,$ the population
difference is oscillating symmetrically between $1$ and $-1$ and its average
is zero. However, for $c/v$ is larger than $2,$ the motion is limited in
half plane and the amplitude decreases with the interaction increasing,
hence, the average of population difference will be nonzero and increasing
with interaction.

\begin{figure}[tbh]
\begin{center}
\rotatebox{0}{\resizebox *{9.0cm}{5.0cm} {\includegraphics
{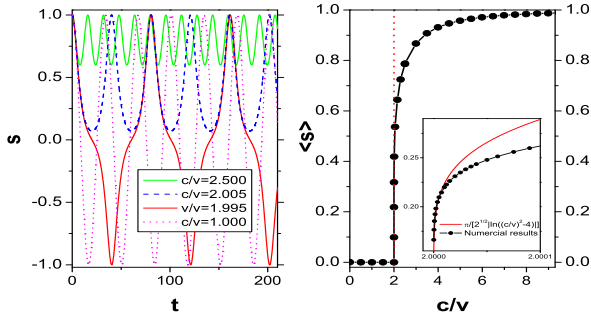}}}
\end{center}
\caption{(Color online) For initial condition $s=1$, the population
difference evolves with time for $c/v=1,1.995,2.005,2.5$ (left column), and
the average of population difference vies parameter $c/v$ (right column).
Inset: the critical behavior near $c/v=2$, the red line is for analytic
formula (5) and the circles are for numerical simulation}
\label{fig.1}
\end{figure}

\begin{figure}[tbh]
\begin{center}
\rotatebox{0}{\resizebox *{9.0cm}{8.0cm} {\includegraphics
{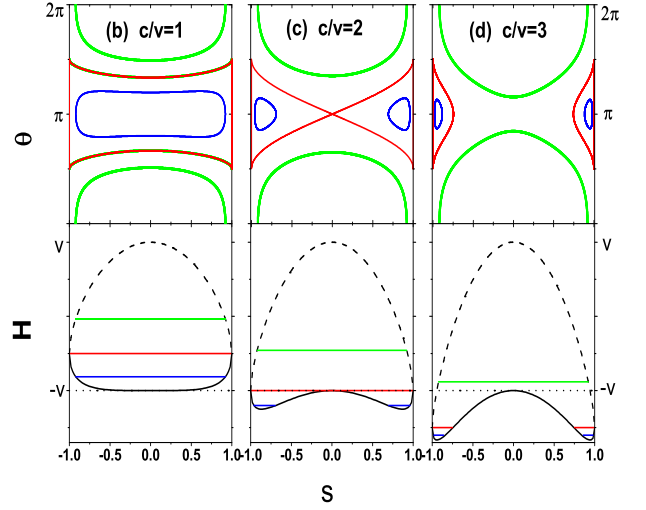}}}
\end{center}
\caption{(Color online) Trajectories on the phase space of the classical
Hamiltonian system (2) (upper panels). In bottom panels we plot the energy
profiles for the relative phase $\protect\theta =0$ (dashed) and $\protect%
\theta =\protect\pi $ (solid), respectively. The energies of the
trajectories in upper panels are also denoted in bottom panels respectively
using the lines with the same colors.}
\label{fig.2}
\end{figure}

The above process can be well understood from the analysis on the phase
space of the classical Hamiltonian system. In Fig.2, we plot the
trajectories in phase space and classical energy profiles for different
parameters. The red lines correspond to the trajectories of which all the
atom are initially in one well, i.e., $s(0)=1$ or $-1.$ From this figure, we
can see clearly that the dynamics transition happens at the moment when the
energy of the trajectory with initial condition $s(0)=1$ or $-1,$ $H(s=\pm
1)=c/2,$ equals to the energy of the separatrix which is $H=-v.$ When the
energy of the trajectory initially in one well is larger than $-v$, one only
finds Josephson oscillation trajectory, when for the energy being smaller
then $-v$, the self-trapping happens.


For classical Hamiltonian system we can obtain the period $T$ of a given
trajectory by the integral $T=\oint \frac{\partial \theta }{\partial H}ds,$
and average $s$ of it by $\left\langle s\right\rangle =\frac{1}{T}\oint
\frac{\partial \theta }{\partial H}sds,$ in which the integral path is along
the trajectory. For the trajectory with initial condition $s(0)=\pm 1,$ we
have $H(s=\pm 1)=c/2.$ Thus, from (\ref{equation3}) and (\ref{eqq}), we get
\begin{equation}
T=\left\{
\begin{array}{cc}
2\int_{-1}^{1}\frac{ds}{v\sqrt{(1-s^{2})-[c(1-s^{2})/2v]^{2}}} & c/v<2 \\
2\int_{\sqrt{1-(2v/c)^{2}}}^{1}\frac{ds}{v\sqrt{(1-s^{2})-[c(1-s^{2})/2v]^{2}%
}} & c/v>2%
\end{array}
,\right.  \label{ttt}
\end{equation}
and
\begin{equation}
\left\langle s\right\rangle =\left\{
\begin{array}{cc}
\frac{2}{T}\int_{-1}^{1}\frac{sds}{v\sqrt{(1-s^{2})-[c(1-s^{2})/2v]^{2}}} &
c/v<2 \\
\frac{2}{T}\int_{\sqrt{1-(2v/c)^{2}}}^{1}\frac{sds}{v\sqrt{%
(1-s^{2})-[c(1-s^{2})/2v]^{2}}} & c/v>2%
\end{array}
,\right.  \label{sss}
\end{equation}
in which we have used the formula $\cos \theta =c\sqrt{1-s^{2}}/2v.$ After
some elaboration, we obtain 
\begin{equation}
\left\langle s\right\rangle =\left\{
\begin{array}{cc}
0 & c/v<2 \\
\frac{\pm \pi \sqrt{(c/v)^{2}-4}}{2c/vIm\{\mathit{K}%
[(c/v)^{2}/((c/v)^{2}-4)]\}} & c/v>2%
\end{array}
,\right.  \label{av}
\end{equation}
where $\mathit{K}(x)$ is the complete first kind of elliptic integral. Near
the transition point, it exhibits the logarithmic critical behavior
\begin{equation}
\left\langle s\right\rangle \approx \pm \frac{\sqrt{2}\pi }{c/v\ln
[(c/v)^{2}-4]}.  \label{power}
\end{equation}
The inset figure of Fig. 1 plots this critical behavior, where theoretical
result is confirmed by numerical result obtained by numerically solving
Eq.(3) with 4th-5th step-adaptive Runge-Kutta algorithm.

Our logarithmic critical behavior is very similar to the critical behavior
in the measure synchronization in coupled Hamiltonian systems \cite{ms}.
This is because critical behavior in both case is closely related to the
separatrix of the Hamiltonian. Near the separatrix the period of the
trajectory diverges to infinity, as the function of the relative deviation
of the energy from the separatrix energy, its divergency follows a logarithm
law\cite{regular}.

In the above discussion, the initial state is set as $s=1$, in fact for any
initial state denoted by $s_i, \theta_i$ the transition to self-trapping
occurs at some interaction parameter, and the critical behavior follows the
same logarithm. If we extend the above discussion to this general case, we
can obtain the general criterion for the occurrence of the self-trapping,
i.e., $H(s_i,\theta_i, c,v) < -v$. Then, the critical point is expressed as,
\begin{equation}
(\frac{c}{v})_{cr}=2(1+\sqrt{1-s_i^2}cos\theta_i)/s_i^2.
\end{equation}
From the above analytic expression, we see that, for the initial state with
smaller population difference it requires stronger nonlinearity so that
self-trapping occurs. Moreover, the critical point can be adjusted by the
relative phase between the two weakly linked BEC in double-well. For
example, for the case the population difference is $0.5$, the critical point
approximates to $15 $ and $8$ for $\theta_i=0$ and $\pi/2$, respectively. In
practical experiments, the relative phase can be adjusted with using
'phase-imprinting' method, i.e., shedding un-uniform laser light on the BECs
in double-well. This method has been successively applied to generate the
dark solitons in cigar-shape BECs\cite{wuliu1}.

\section{Many Body Quantum Fluctuation Effects}

In the mean field treatment we assume that the number of particle is
large enough. However, in practical experiment, the particle number
is finite, in order to know the quantum fluctuation effect due to
finite particle number, we should investigate the self-trapping
within the framework of the many body quantum system
(\ref{ham})\cite{smezir,mlb}.

In treating the quantum many-body problem it is helpful to bear in mind some
results from the quantum information theory concerning the entanglement \cite%
{mmm}. The quantum entanglement is realized not only to be a crucial
resource that allows for powerful communication and computational tasks that
are not possible classically, but also to be a signal for quantum long-range
correlation and therefore can serve as indicator for the quantum transition
in concrete solid system\cite{addadd1}. Recent years witness growing
interests in studying the interplay between entanglement and quantum phase
transition\cite{ccc,addadd2,lhq,addsolid,addadd3}.

Previously, some efforts have been devoted to study the dynamics
of the BECs in double-well with full quantum treatment
\cite{pradd,jpadd}. In Ref.\cite{pradd} the authors presented a
quantum phase-space model of the BECs in a double-well potential
by using the Husimi distribution function. They showed a good
correspondence between the phase space of classical Hamiltonian
(\ref{equation3}) (mean-field approximation) and the quantum
phase-space of two-mode hamiltonian (\ref{ham}) (full quantum
framework). The authors of Ref. \cite{jpadd} calculated the time
evolutions of states and their corresponding entanglement with
different initial states for several different interactions
between atoms. The time evolutions of entanglement entropy
presented in Ref.\cite{jpadd} for several interactions between
atoms show a decreasing tendency with increasing interactions.

In our following discussions of this section, however,  we focus
on the critical behavior at the transition to self-trapping as
revealed by the above discussions, addressing how the quantum
fluctuation influence on the transition behavior. As will be shown
latter, the transition point is shifted and the scaling-law is
broken down due to the quantum fluctuation. With increasing the
atom number, the transition behavior demonstrates a perfect
classical quantum correspondence. We also calculating the
entanglement entropy achieving insight into the quantum
transition. Our calculations on the time evolutions of
entanglement entropy confirm the results of \cite{jpadd}, and our
further calculations on the time averaged entanglement strongly
suggest that the entanglement entropy of this system serve as  a
good order parameter to describe such quantum transition.

\subsection{Quantum Phase Transition}

In the quantum framework the evolution of the system is governed by the Schr%
\"{o}dinger equation
\begin{equation}
i\frac{d}{dt}\left| \psi (t)\right\rangle =\widehat{H}\left| \psi
(t)\right\rangle ,  \label{sc}
\end{equation}
where $\left| \psi (t)\right\rangle =\sum_{n=0}^{N}a_{n}\left|
n,N-n\right\rangle $, $\left| n,N-n\right\rangle =\frac{1}{\sqrt{n!(N-n)!}}%
\left( \widehat{a}^{\dagger }\right) ^{n}\left( \widehat{b}^{\dagger
}\right) ^{N-n}\left| 0\right\rangle $ ($n=0,\cdots ,N$) are Fock states,
and $a_{n}$ are the probability amplitudes respectively. Hence, the
population difference is given by
\begin{equation}
s=\sum \frac{\left| a_{n}\right| ^{2}\left( N-2n\right) }{N}.  \label{ss}
\end{equation}

We choose $\left| 0,N\right\rangle $ as the initial state in the full
quantum framework, which corresponds to $s=1\ $in the mean field model. In
Fig. 3, we plot the average population difference calculated from the above
Schr\"{o}dinger equation for different total particle numbers. From the
above calculation, we find that, the quantum fluctuation has two significant
effects on the transition to self-trapping. First, the critical point is
shifted to the left-hand side due to the finiteness of the particle number.
From Fig.3, it is clearly seen that, keeping parameters $c,v$ as constant,
with increasing the number of the atoms, the transition point clearly shift
to left-hand. For example, for $N=50$, the transition point shifts to $%
c/v=1.6$. We know that the quantum fluctuation closely relates to effective
planck constant, here for this model it is $\frac{2}{N}$ \cite{wuliu}.
Therefore we expect the deviation from the mean field critical point should
be inversely proportional to total atom number. This prediction is confirmed
by our calculations as shown in the inset of Fig.3. in which $\delta $ is
the difference between the quantum transition point and the mean field one.

Secondly, the logarithm scaling-law is broken down by the quantum
fluctuation. In Fig.3 we find the quantum fluctuation destroys the logarithm
scaling-law of the mean field and no clear scaling-law is observed for the
quantum case. With increasing the atom number the quantum results tends to
the mean field results in the limit $N$ to $\infty $, as required by the
classical quantum correspondence principle\cite{wuliu}.

We should address that in the above calculation time period for average
should be much longer than the period of the fastest oscillations but
shorter than the period of the shortest quantum beating. This is because,
essentially, the dynamics of quantum system is periodic or quasi-periodic,
therefore, any dynamical effects of quantum system depends on the time
scales \cite{s24,smezir}. In our problem, there are two time scale, one is
for integrating classical equation (3), the other is for integrating quantum
equation (10), the later is $N$ times the former one. In our calculations,
the average time is $50 N$, meaning that the corresponding classical time
scale is same (=50). On the other hand, we find that,the averaged population
difference increases gradually to $0.001$ and then soars up. This is
different from the mean field situation, where the averaged population
difference keeps zero and then turns to be nonzero after critical point. So,
in the quantum case, we define the transition point as the point that the
averaged population difference is larger than $0.001$. This observation
also suggests that the no scaling-law for the finite particle  situation.

\begin{figure}[tbh]
\begin{center}
\rotatebox{0}{\resizebox *{9.0cm}{8.0cm} {\includegraphics
{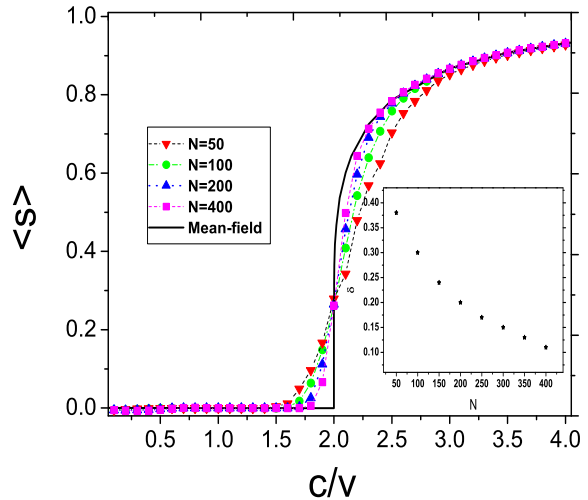}}}
\end{center}
\caption{(Color online) The average of population difference vies the
parameter $c/v$ obtained by full quantum simulations. Here, the initial
state is $(n=0,\cdots ,N)$, namely $s=1$. Insert: the shift of the critical
point comparing with meanfield predication $\protect\delta $ for different
atom number. }
\label{fig3}
\end{figure}

\subsection{Entanglement Manifestation of the Quantum Phase Transition}

To well understand the self-trapping phenomenon in full quantum description,
we calculate evolution of the occupations on Fock states $\left|
n,N-n\right\rangle $ $(n=0,\cdots ,N)$. Fig. 4 shows the evolution of Fock
state occupation for different interactions. The horizontal axis is the time
$t$, the vertical axis is the index of Fock state, namely, $n$ corresponds
to $\left| n,N-n\right\rangle $, and the contour is for the occupation
probability. Fig. 4(a) is for the linear case, which shows the occupations
are oscillating between $\left| 0,N\right\rangle $ to $\left|
N,0\right\rangle .$ For $c=1.95$ (seeing Fig.4(b), near the transition
point, we see that the wavefunction spread much rapidly to all the Fock
states. Fig. 4(c) is plotted for the self-trapping case, $c=2.5,$ from which
we see that the occupations is narrowed\ in partial Fock states, so that the
average of population difference is nonzero.

\begin{figure}[!tbh]
\begin{center}
\rotatebox{0}{\resizebox *{9.0cm}{8.0cm} {\includegraphics
{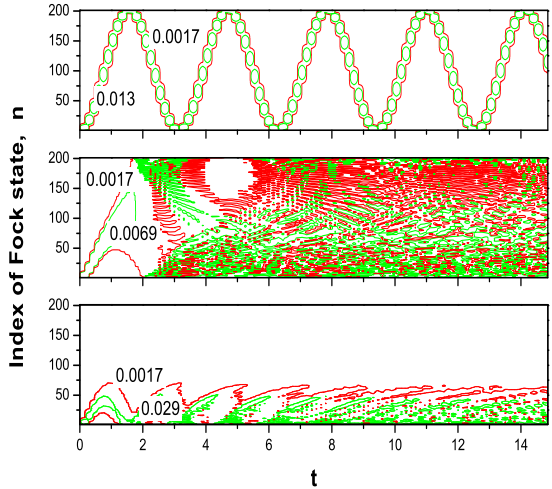}}}
\end{center}
\caption{(Color online) Occupations on Fock states for different time. The
contours denote the probabilities of occupations. From up to bottom, $c/v=0,
1.995, 2.5$ respectively.}
\label{fig4}
\end{figure}

From Fig. 4, we also see that the dynamics properties of such a quantum
system are quite different for different interaction regions. To achieve
more insight into the quantum transition to the self-trapping, we introduce
the quantum entanglement entropy. For the system with the wave function $%
\left| \psi \right\rangle =\sum_{n=0}^{N}a_{n}\left| n,N-n\right\rangle $,
its density operator is given by
\begin{equation}
\rho =\left| \psi \right\rangle \left\langle \psi \right|
=\sum_{n,m}a_{n}a_{m}^{\ast }\left| n,N-n\right\rangle \left\langle
N-m,m\right| ,  \label{density}
\end{equation}
Taking the partial trace with respect to one well yields the reduced density
operator for the other,
\begin{equation}
\rho _{a}=\sum_{n}|a_{n}|^{2}\left| n\right\rangle \left\langle n\right| .
\label{densa}
\end{equation}
Thus, the entropy of entanglement between the two coupled BEC's is given by %
\cite{mlb}
\begin{equation}
E(\rho )=-\sum_{n=0}^{N}|a_{n}|^{2}\log |a_{n}|^{2}.  \label{entr}
\end{equation}
The entanglement entropy has the following properties: its reaches maximum $%
E(\rho )=\log N$ when $|a_{n}|^{2}=\frac{1}{N},$ and its minimum $E(\rho )=0$
when $|a_{n}|^{2}=1$ and others are zero.

\begin{figure}[!tbh]
\begin{center}
\rotatebox{0}{\resizebox *{9.0cm}{8.0cm} {\includegraphics
{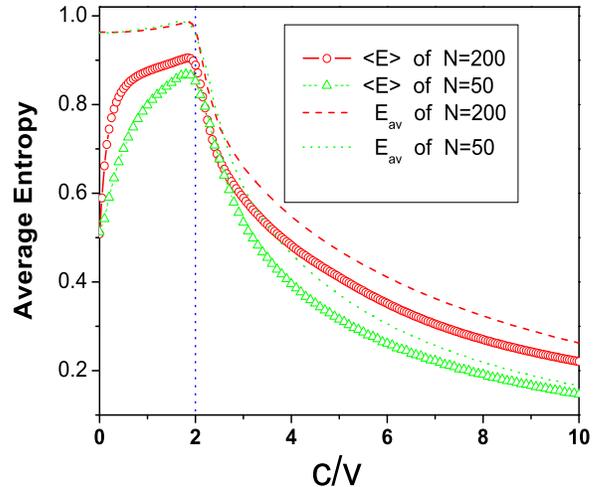}}}
\end{center}
\caption{(Color online) The entanglement entropy for different $c/v$. The
circle line is for $E_{av}$ and dashed line for $\left\langle E\right\rangle$%
}
\label{fig5}
\end{figure}

Because the self-trapping is dynamic phenomenon, the occupation on each Fock
state is varied in time, therefore, we use average entropy. Technically, we
have two choices to average it: we can average occupation firstly and then
calculate entropy, and calculate entropy firstly then average it. So, we
denote
\begin{equation}
E_{av}=-\sum_{n=0}^{N}\left\langle |a_{n}|^{2}\right\rangle \log
\left\langle |a_{n}|^{2}\right\rangle /\log N,  \label{ave}
\end{equation}
and
\begin{equation}
\left\langle E\right\rangle =\left\langle -\sum_{n=0}^{N}|a_{n}|^{2}\log
|a_{n}|^{2}\right\rangle /\log N.  \label{eve}
\end{equation}
The above formula have been normalized by $\log N.$

In Fig. 5, we plot the two kinds of average entropy with different
interaction. Obviously, the dynamics of the quantum system can be well
illustrated by the average entropies in quantities. For the linear case, the
atoms mainly occupy several Fock states at a given time (see Fig. 4 (a)),
and the occupied states are changed with time. Thus, the instantaneous
entropy is small, so does the average entropy $\left\langle E\right\rangle .$
On the other hand, because the average population on each Fock state are
almost equivalent for this case, so $E_{av}$ should be large. With the
interaction increasing, the occupations on Fock states extend so that the
instantaneous entropy increases. However, when the interaction exceeds the
transition point self-trapping occurs and the occupations are limited on
several Fock states (see Fig.4 (c)), hence, the instantaneous entropy
becomes small, so does $\left\langle E\right\rangle .$ For the same reason $%
E_{av}$ will be small for self-trapping cases. From Fig. 5, we
also observe that $E_{av}$ is almost independence of the particle
number and varies very little with changing the interaction
parameter before it reaches its maximum. After that, it shows
quite sensitive on the particle number as well as the interaction
parameter. Whereas, for the entropy $<E>$ before and after the
maximum point it shows strong dependence of the particle number.

It is interesting that the two average entropies reach their
maximum at the point very close to the transition point of mean
field $c/v=2$. This is very similar to phase transition of spin
systems, where the phase transition happens at the point when
entanglement of system reaches maximum \cite{lhq}. This is
different to the situation of the two-impurity Kando model where
the entanglement vanishes at a quantum critical
point\cite{addsolid}. In particular, we find that, the maximum
points of the average entropies varies very little with changing
the particle number. The property suggests that the entanglement
entropy is a better quantity than the average population in
serving as a indicator to quantum phase transition, because the
latter is too sensitive on the particle number as shown in Fig.3.

\section{Discussion and Conclusion}

In conclusion, we have made thoroughly analysis on the transition to
self-trapping for BECs confined in a symmetric double-well. Analytically we
identify it as a continuous phase transition, where the time averaged
population difference between two wells changes from zero to nonzero
following a logarithm law at a critical point. We also discuss influence of
the many-body quantum fluctuation on the transition to self-trapping. We
find that the transition point is significantly shifted by quantum
fluctuation and no scaling-law is observed in quantum description. We
investigate the quantum entanglement manifestation of the transition and
find that the entangle entropy reaches its maximum at the transition point.
Classical quantum correspondence in the transition process is discussed.

\section{Acknowledgments}

This work was supported by National Natural Science Foundation of
China (No.10474008,10604009), Science and Technology fund of CAEP,
the National Fundamental Research Programme of China under Grant
No. 2005CB3724503, the National High Technology Research and
Development Program of China (863 Program) international
cooperation program under Grant No.2004AA1Z1220. We thank Y.Ma for
her help in some calculations.


\begin{thebibliography}{*}
\bibitem[*]{} Liu$\_$Jie@iapcm.ac.cn

\bibitem{grifoni} M. Grifoni, P. H\"{a}nggi, Phys. Rep. 304, 229(1998)

\bibitem{anderson} M. H. Anderson, J. R. Ensher, M. R. Matthews, C. E.
Wieman, and E. A. Cornell, Science 269, 198 (1995); K. B. Davis, M. -O.
Mewes, M. R. Andrews, N. J. van Druten, D. S. Durfee, D. M. Kurn, and W.
Ketterle, Phys. Rev. Lett. 75, 3969 (1995); C. C. Bradley, C. A. Sackett, J.
J. Tollett, and R. G. Hulet, ibid. 75, 1687(1995)

\bibitem{franco} Franco Dalfovo, Stefano Giorgini, L.P. Pitaevskii, S.
Stringari, Rev. Mod. Phys. 71, 463(1999); Anthony J. Leggett, Rev. Mod.
Phys. 73, 307(2001), and references therein.

\bibitem{milburn} O.Zobay and B.M.Garraway, Phys. Rev. A. \textbf{61},
033603(2000); Biao Wu and Qian Niu, Phys. Rev. A 61, 023402(2000); F.Kh.
Abdullaev and R.A.Kraenkel, Phys. Rev. A \textbf{62}, 023613 (2000); F.Meier
and W.Zwerger, Phys. Rev. A \textbf{64}, 033610 (2001).

\bibitem{JieLiu} Jie Liu, Biao Wu and Qian Niu, Phys. Rev. Lett. \textbf{90}%
, 170404(2003); Jie Liu et. al. Phys. Rev. A \textbf{66}, 023404(2002).

\bibitem{gio} Giovanazzi et al., Phys. Rev. Lett. \textbf{84} (2000) 4521;
Li-Bin Fu, Jie Liu and Shi Gang Chen, Phys. Lett. A \textbf{298}, 388(2002).

\bibitem{shin} Y.Shin, M. Saba, T. A. Pasquini, W. Ketterle, D. E.
Pritchard, and A. E. Leanhardt, Phys. Rev. Lett. \textbf{92}, 050405 (2004)

\bibitem{michael} M. Albiez, R. Gati, Jonas F\"olling, S. Hunsmann, M.
Cristiani, and M.K. Oberthaler, Phys. Rev. Lett. \textbf{95}, 010402(2005)

\bibitem{selftrap} G. J. Milburn et. al., Phys. Rev. A \textbf{55},
4318(1997);

\bibitem{smerzi} A. Smerzi, S. Fantoni, S. Giovanazzi and S.R. Shenoy, Phys.
Rev. Lett. \textbf{79}, 4950(1997).

\bibitem{raghavan} S. Raghavan, A. Smerzi and V.M. Kenkre, Phys. Rev. A
\textbf{60}, R1787(1999)

\bibitem{raghavan1} S. Raghavan, A. Smerzi, S. Fantoni and S. R. Shenoy,
Phys. Rev. A \textbf{59}, 620(1999)

\bibitem{sigmund} Sigmund Kohler and Fernando Sols, Phys. Rev. Lett. 89,
060403(2002); L. Salasnich, Phys. Rev. A \textbf{61}, 015601 (2000);

\bibitem{s14} M.J. Steel and M.J. Collett, Phys. Rev. A \textbf{57}, 2920
(1998); J.I. Cirac, M. Lewenstein, K. M\O mer, and P. Zoller, Phys. Rev. A
\textbf{57}, 1208(1998).

\bibitem{ms} A. Hampton and D.H. Zanette, Phys. Rev. Lett. \textbf{83},
2179(1999).

\bibitem{regular} A.J.Lichtenberg and M.A.Lieberman, 'Regular and Chaotic
Dynamics', p236, Springer-Verlag (Second edition) (1983).

\bibitem{wuliu1} Biao Wu, Jie Liu, and Qian Niu Phys. Rev. Lett. \textbf{88}%
, 034101 (2002).

\bibitem{smezir} S. Raghavan, A. Smerzi, and V.M. Kenkre, Phys. Rev. A
\textbf{60}, R1787(1999).

\bibitem{mlb} Andrew P. Hines, Ross H. McKenzie, and Gerard J. Milburn,
Phys. Rev. A \textbf{67}, 013609(2003).

\bibitem{mmm} A. Galindo and M.A. Martin-Delgado, Rev.Mod.Phys. 74
347-423(2002).

\bibitem{addadd1} T.J.Osborne and M.A.Nielsen, Phys. Rev. A \textbf{66},
032120 (2002); A.Osterloh, L.Amico, G.Falci, and R.Fazio, Nature (London)
\textbf{416}, 608 (2002).

\bibitem{ccc} F. Verstraete, M.A. Martin-Delgado, J.I. Cirac, Phys. Rev.
Lett. 92, 087201 (2004); J. J. Garcia-Ripoll, M. A. Martin-Delgado, J. I.
Cirac,  Phys. Rev. Lett. 93, 250405 (2004).

\bibitem{addadd2} J.Vidal, R.Mosseri, and J.Dukelsky, Phys.Rev.A \textbf{69}
054101(2004); J.Vidal, Guillaume Palacios, and Claude Aslangul Phys. Rev. A
\textbf{70} 062304 (2004).

\bibitem{lhq} Shi-Jian Gu, Shu-Sa Deng, You-Quan Li, and Hai-Qing Lin, Phys.
Rev. Lett. \textbf{93}, 086402(2004); Shi-Jian Gu, Guang-Shan Tian, and
Hai-Qing Lin, Phys. Rev. A \textbf{71}, 052322 (2005).

\bibitem{addsolid} C.Brukner, V.Vedral and A.Zeilinger, Phys. Rev. A \textbf{%
73}, 012110 (2006); S.Y.Cho and R.H.Mckenzie, Phys. Rev. A \textbf{73},
012109 (2006).

\bibitem{addadd3} Yan Chen, Z.D.Wang, and F.C.Zhang, Phys. Rev. B \textbf{73}
224414(2006)



\bibitem{pradd} K. W. Mahmud, H. Perry, and W. P. Reinhardt, Phys. Rev. A 71, 023615 (2005).

\bibitem{jpadd} A. P. Tonel, J. Links, and A. Foerster, J. Phys. A: Math. Gen. 38, 1235--1245
(2005).

\bibitem{wuliu} Biao Wu and Jie Liu, Phys. Rev. Lett. \textbf{96},
020405(2006).

\bibitem{s24} V.M. Kenkre, M.F. J$\phi $rgensen, and P.L. Christiansen,
Physica D \textbf{90}, 280 (1996).

\end{thebibliography}
\end{document}